\documentclass[sigconf]{acmart}

\AtBeginDocument{%
  }

\setcopyright{acmlicensed}
\copyrightyear{2024}
\acmYear{2024}
\acmDOI{XXXXXXX.XXXXXXX}

\acmConference[APNet'24]{$8^{th}$ Asia-Pacific Workshop on Networking}{August 3-4, 2024}{Sydney, Australia}
\acmISBN{978-1-4503-XXXX-X/18/06}

\begin{document}
\copyrightyear{2024}
\acmYear{2024}
\setcopyright{rightsretained}
\acmConference[APNet 2024]{The 8th Asia-Pacific Workshop on Networking}{August 3--4, 2024}{Sydney, Australia}
\acmBooktitle{The 8th Asia-Pacific Workshop on Networking (APNet 2024), August 3--4, 2024, Sydney, Australia}\acmDOI{10.1145/3663408.3665822}
\acmISBN{979-8-4007-1758-1/24/08}
\title{Optical-computing-enabled Network: A New Dawn for Optical-layer Intelligence?}

\author{Dao Thanh Hai}
\affiliation{%
  \institution{RMIT University Vietnam}
  \city{Ho Chi Minh}
  \country{Vietnam}
}
\email{hai.dao5@rmit.edu.vn}
%\author{Minh Nguyen}
%\affiliation{%
  %\institution{RMIT University Vietnam}
  %\city{Ho Chi Minh}
  %\country{Vietnam}

%\email{minh.nguyen3@rmit.edu.vn}
\author{Minh Nguyen}
\affiliation{%
  \institution{RMIT University Vietnam}
  \city{Ho Chi Minh}
  \country{Vietnam}
}
\email{minh.nguyen326@rmit.edu.vn}

\author{Isaac Woungang}
\affiliation{%
  \institution{Toronto Metropolitan University}
  \city{Toronto}
  \country{Canada}
}
\email{iwoungan@torontomu.ca}

%\author{Fen Zhou}
%\affiliation{%
  %\institution{University of Avignon}
  %\city{Avignon}
  %\country{France}
%}
%\email{fen.zhou@univ-avignon.fr}


\begin{abstract}
Inspired by the renaissance of optical computing recently, this poster presents a disruptive outlook on the possibility of seamless integration between optical communications and optical computing infrastructures, paving the way for achieving optical-layer intelligence and consequently boosting the capacity efficiency. This entails a paradigm shift in optical node architecture from the currently used optical-bypass to a novel one, entitled, optical-computing-enabled mode, where in addition to the traditional add-drop and cross-connect functionalities, optical nodes are upgraded to account for optical-computing capabilities between the lightpath entities directly at the optical layer. A preliminary study focusing on the optical aggregation operation is examined and early simulation results indicate a promising spectral saving enabled by the optical-computing-enabled mode compared with the optical-bypass one.   
\end{abstract}



\maketitle

\section{Introduction}
Optical communications and networks underpinning the global Internet and digital infrastructure have been evolving technologically and architecturally to meet the explosively growing traffic demands. While technology-based breakthroughs expand the capacity by order-of-magnitude, network architecture innovations aim instead at harnessing the available capacity in an optimal manner so that more traffic could be supported at less spectrum and/or resources, resulting in greater operational efficiency. However, it appears that optical network architecture has remained essentially unchanged since the arrival of optical-bypass around the year 2000s and then has become widely adopted nowadays by worldwide network carriers \cite{Hai2023}. In optical-bypass mode, the common wisdom lies in ensuring a lightpath optically intact from the source to the destination and in doing so, in-transit lightpaths that are routed over a common intermediate node must be separated with each other either in time, frequency or spatial domain in avoidance of unwanted interference. This poses a critical bottleneck for improving the flexibility and scalability, particularly in view of the resurgences in optical computing technologies, permitting various optical mixing operations between such lightpaths to compute new ones of potentially greater capacity-efficiency \cite{Zhong2023}. \\
\indent{We envision a disruptive architecture for next-generation optical transport networks, i.e., featuring the integration of optical computing functionalities into the optical layer, potentially marking the dawn of the optical-layer intelligence where the lightpaths carrying wavelength-channel rate (e.g., Tb/s) could be optically interfered to each other (i.e., analog computing) to produce more efficient ones}. Our proposal, named, optical-computing-enabled network thus entails a paradigm shift in optical node architecture, where in addition to the traditional add-drop and cross-connect functionalities as currently implemented in optical-bypass mode, optical nodes are upgraded to include optical-computing capabilities between lightpath entities. In this paper we are interested in the optical (de-) aggregation operations whose enabling technologies have been progressively maturing \cite{apl}. In the next Section, we illustrate how optical-computing-enabled network utilizing optical aggregation could be spectrally more efficient than the optical-bypass mode in supporting traffic demands. 
\section{Optical-computing-enabled Network vs. Optical-bypass Network}
Supposing that there are two 400G requests from node $A$ and node $B$, both to node $C$. In optical-bypass mode, two lightpaths with appropriate routes and wavelengths would be set up and one such traffic provisioning is shown in Fig. 1(a). As the two lightpaths are routed over the same link $XC$, two distinct wavelengths must be available on the link $XC$. The spectral cost is thus two wavelength count and four wavelength-link units. In the new paradigm of optical-computing-enabled mode, in which in-network optical computing is available at the optical nodes, new opportunities emerge. Specifically, we focus on the case that the all-optical aggregation capability is empowered at node $X$. Under this assumption, a new integrated lightpath modulated on the higher-order format 16-QAM with the wavelength $\lambda_1$ carrying the traffic of both demands could be created at node $X$ by adding together two lightpaths $A_{\lambda_1}$ and $B_{\lambda_1}$ (Fig. 1(b)). At the shared destination node $C$, the aggregated lightpath would undergo the reverse process, i.e., an optical de-aggregation to recover the individual original ones (Fig. 1(c)). The spectral cost for this approach would be one wavelength count (i.e., $\lambda_1$) and three wavelength-link units, suggesting a greater efficiency in harnessing the optical spectrum. In this case, the computing sense is interpreted as the addition of bits per symbol from two ordinary lightpaths (i.e., QPSK: 2 bits/symbol) to a new integrated one (i.e., 16-QAM: 4 bits/symbol as in Fig. 1(c)). Note that the optical-computing-enabled paradigm promises greater networking flexibility thanks to the possibility of mixing the lightpath entities for computing purposes, thus posing further challenges for the network design algorithms to tap into the potential benefits. 
	\begin{figure}[!ht]
		\centering
		\includegraphics[width=\linewidth, height = 6.5cm]{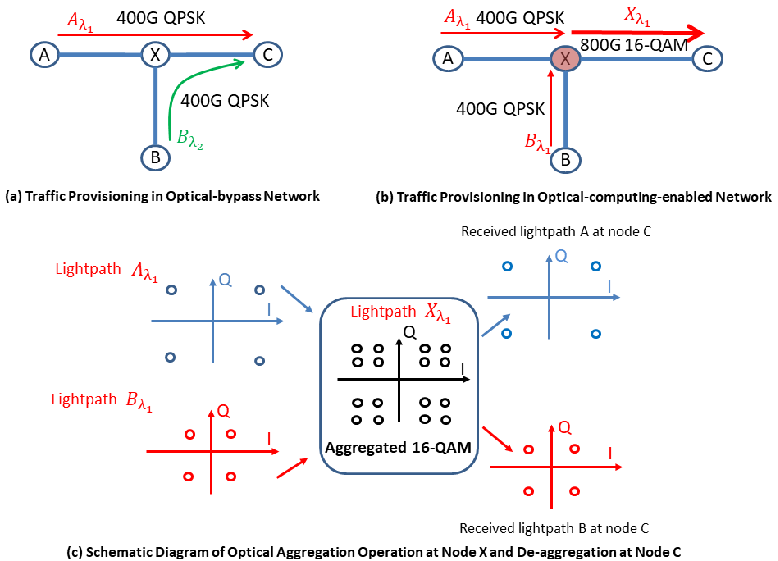}
		\caption{Traffic provisioning in optical-computing-enabled mode with optical (de-)aggregation}
		\label{fig:i1}
	\end{figure}
\section{Preliminary Evaluations}
To evaluate the spectral efficiency of optical-computing-enabled network with optical aggregation operation, we formulate in the form of an integer linear programming (ILP) model the newly arisen problem, entitled, \textit{routing, wavelength and computing assignment}, where the new dimension of computing involves the optimal determination of pair of demands for computing, the nodes at which the computing is performed and the routing and wavelength / spectrum for such computed lightpaths. Compared to the conventional \textit{routing and wavelength assigment} in optical-bypass network, such new problem is one order of magnitude computationally harder. Two realistic network topologies, NSFNET and INDIA, with two-to-all traffic model are investigated (Fig. 2(a), (b), (c)). Ten traffic samples are generated and all the results are optimally collected from solving the developed ILP model to guarantee a reliable comparison. As it can be observed from Fig. 2(d), the maximum relative gain for both two networks is around $35\%$ while the average gain is more than $16\%$ and the denser the topology is, the more likely it could spectrally benefit from exploiting the optical-computing-enabled mode. These preliminary results are provided to demonstrate the feasibility of our developed mathematical network design model, which has been evaluated on realistic networks. The readers then can use our illustration and make such comparison for other network topologies and traffic models that are of interest to them. It has to be noted that we have developed the mathematical design framework that could be applicable to any traffic model and any topology. The precise quantitative conclusions about which the network performs better and exactly by how much in terms of spectrum and resources savings are limited to the utilized network and the traffic type we use. \\
\begin{figure}[!ht]
		\centering
		\includegraphics[width=\linewidth, height = 7cm]{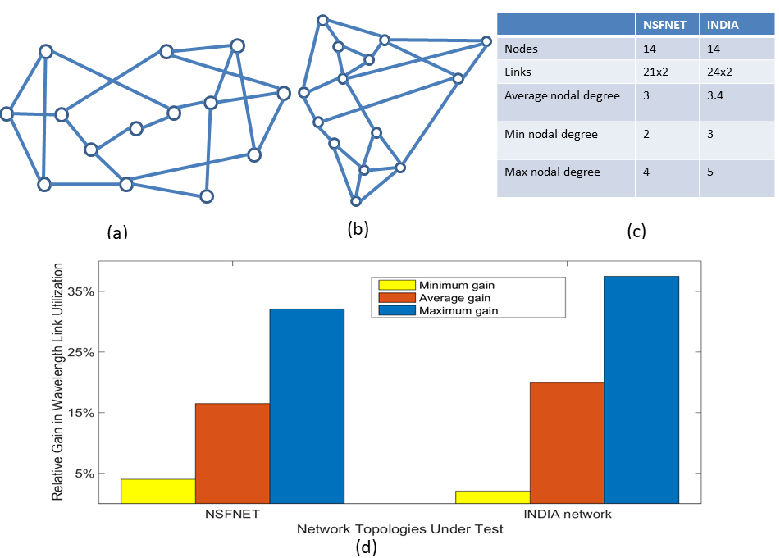}
		\caption{a) NSFNET topology b) INDIA network c) Network topologies characteristics d) Relative gain in wavelength link utilization between optical-computing-enabled mode versus optical-bypass one}
		\label{fig:i1}
	\end{figure}
 
\section{Conclusions}
With the growing maturing of optical computing platforms, we envision a transition from the optical-bypass architecture to optical-computing-enabled one, where optical computing and communications could be seamlessly integrated at the optical layer, marking the dawn of optical-layer intelligence going beyond its traditional high-speed connectivity function.   

\bibliographystyle{ACM-Reference-Format}
\bibliography{apnet}

\end{document}